\documentclass[epj]{svjour}

\usepackage{graphics}
\usepackage{fancyhdr}
\usepackage{amssymb}
\usepackage{epsfig}

\def\mytitle{My title}
\def\myauthors{My name}
\def\mytype{My type of session}
\def\mysession{My session}
\def\mytitle{SUSY Model Building} %Put your title here!
\def\myauthors{Stuart Raby}    %Put your name here!
\def\mytype{Review}
\def\mysession{\myauthors}
\begin{document}
\title{SUSY Model Building}

\author{Stuart Raby
\thanks{\emph{ Email:} raby@pacific.mps.ohio-state.edu, \newline OHSTPY-HEP-T-07-004} }
 \institute{The Ohio State University, 191 W. Woodruff Ave,
Columbus, OH 43210, USA}

\date{}
\abstract{ I review some of the latest directions in supersymmetric
model building, focusing on SUSY breaking mechanisms in the minimal
supersymmetric standard model [MSSM], the ``little" hierarchy and
$\mu$ problems, etc.  I then discuss SUSY GUTs and UV completions in
string theory.
\PACS{
      {PACS-key}{discribing text of that key}   \and
      {PACS-key}{discribing text of that key}
     } % end of PACS codes
} %end of abstract
\maketitle
\section{Introduction}
\label{intro} The Supersymmetric Standard Model is motivated to
solve the gauge hierarchy problem, i.e. to explain the small number
$M_Z/M_{Pl} \sim 10^{-16}$.   In the Standard Model this requires a
fine-tuning of one part in $10^{32}$.   Supersymmetry makes this
gauge hierarchy ``technically natural" since scalar masses are tied
to the value of their fermionic partners.   And fermions have chiral
symmetries which keep their masses only logarithmically sensitive to
UV physics.  Finally, if SUSY is spontaneously broken by some
dynamical mechanism, we obtain a SUSY breaking scale $M_{SUSY}$ of
order $\approx e^{-\frac{8\pi^2}{g^2(M_{Pl})}} M_{Pl}$.

On the other hand, SUSY GUTs are motivated by the quest for
understanding -
\begin{itemize}
\item  charge quantization;
\item  family structure;
\item  gauge coupling unification,
which works and defines a new scale of nature at $M_G \sim 10^{16}$
GeV,  with  $M_{SUSY} \sim 1$ TeV; and
\item neutrino masses,
with a See-Saw scale of order $(0.01 - 0.1) \times M_G$.
\end{itemize}
\subsection{Minimal Supersymmetric Standard Model}
The MSSM is defined by its minimal spectrum defined by the
superfields -
\begin{itemize}
\item $Q, U^c, D^c, L, E^c, N^c$ - fermions
and sfermions;
\item $V^i, \ (i= 1,2,3)$ - gauge bosons
and gauginos;
\item  $H_u, \ H_d $ - Higgs and
Higgsinos.
\end{itemize}

And a $\mathbb{Z}_2$ symmetry defined by the operation -
$$ F \longrightarrow - F, \;\; H \longrightarrow H $$
where $F, \ H$ are matter and Higgs multiplets, respectively.  This
symmetry is sometimes called R-parity (if it distinguishes particles
and their superpartners), or family reflection symmetry  (matter
parity) (if it does not distinguish particle and superpartner). This
symmetry forbids the dangerous dimension 3 and 4 baryon and lepton
number violating operators and guarantees that the lightest
superpartner is stable and is a possible dark matter candidate.
\subsection{SUSY Model Building}
Perhaps before discussing the recent progress in SUSY model
building, it may be worthwhile to put this whole program in context.
There are several well-known problems,  including the SUSY flavor
problem,  the $\mu$ problem,  the ``little" hierarchy problem, the
SUSY CP problem,  and the grand unified symmetry breaking  and
doublet-triplet splitting problems.   In the first stage of model
building, one searches for ``mechanisms" which can solve these
problems.   Typically two or more ``mechanisms" for solving a
certain set of problems are mutually exclusive; meaning you have
solved only one of these problems in the set and you can choose
which problem you want to solve.   It is of course much better to
find one ``mechanism" which solves more than one problem or a
self-consistent set of ``mechanisms" which simultaneously solve a
set of problems.  Finally, in the best case one would be able to
solve all known problems in one self-consistent theory.   This is
clearly the goal of SUSY model building.   With that said, let us
now discuss some recent progress in SUSY model building.

\section{Dynamical SUSY Breaking}
There has been recent work on the subject of dynamical SUSY breaking
in global supersymmetric theories by the following long list of
authors - Intriligator, Seiberg \& Shih; Dine, Feng \& Silverstein;
Kitano, Ooguri \& Ookouchi; Argurio, Bertolini, Franco \& Kachru;
Murayama \& Nomura; Dine \& Mason; Brummer; Bai, Fan \& Han; Dudas,
Mourad \& Nitti; Gomes-Reino \& Scrucca; Amariti, Girardello \&
Mariotto; Essig, Sinha \& Torroba; Ahn; Serone \& Westphal; Cho \&
Park; Abel, Dumford, Jaeckel \& Khoze; Tatar \& Wetenhall; van den
Broek; Ferretti; Pastras; Ooguri, Ookouchi \& Park; Kawano, Ooguri
\& Ookouchi.   The bottom line of this work is that meta-stable
SUSY breaking vacua in {\em global} SUSY are quite probable!   In
fact, in the talk by Murayama (this conference) it was argued that
dynamical SUSY breaking is, in fact, no longer an obstacle to model
building.

On the other hand,  in another parallel series of recent articles by
- Kachru, Kallosh, Linde \& Trivedi; Choi, Falkowski, Nilles,
Olechowski, Pokorski; Endo, Yamaguchi \& Yoshioka; Choi, Jeong \&
Okumura;  \newline Falkowski, Lebedev \& Mambrini; Kitano \& Nomura;
Lebedev, Nilles \& Ratz; Lebedev, Loewen, Mambrini, Nilles \& Ratz;
Acharya, Bobkov, Kane, Kumar \& Vaman (Shao); Randall \& Sundrum;
Giudice, Luty, Murayama \& Rattazzi -  it has been argued that
meta-stable SUSY breaking vacua are generic in {\em local} SUSY
/string theory vacua!

The bottom line of this general analysis is that meta-stable SUSY
breaking vacua are ubiquitous and they have cosmologically long
life-times. Of course, the latter feature is a very satisfying
prerequisite.

I do not have time to elaborate in this talk on this general and
more formal aspect of SUSY model building.   Instead I will now
discuss some specific examples of SUSY breaking mechanisms within
the context of solving some of the SUSY problems mentioned earlier.
The most important aspect of SUSY breaking relevant for low energy
phenomenology concerns the mechanism for transmitting SUSY breaking
from the {\em SUSY breaking sector} to the {\em observable sector}
of the theory.   There are still three fundamental mechanisms for
mediating SUSY breaking known as -
\begin{itemize}
\item gravity mediated SUSY breaking,
\item gauge mediated SUSY breaking, and
\item anomaly mediated SUSY breaking.
\end{itemize}
In addition there are also variations of the above, known as - eg.
gaugino, moduli, and dilaton mediation.

\section{Focus on solving problems of the MSSM}

\subsection{``Little" hierarchy problem}

LEP II data excludes a Standard Model Higgs boson with mass less
than 114.4 GeV.   This lower bound has generated a great deal of
angst among SUSY aficionados.  The reason is that, as we will now
argue, this requires a fine-tuning of parameters.   This is known as
the ``little" hierarchy problem.

Consider the one loop corrected value of the Higgs mass in the MSSM.
We have
\begin{eqnarray}
m_h^2 \approx M_Z^2 + \frac{3 G_F m_t^4}{\sqrt{2} \pi^2} \left( \log
\frac{m_{\tilde t}^2}{m_t^2} + X_t^2 (1 - \frac{X_t^2}{12}) \right)
\end{eqnarray}
where $X_t^2 = |A_t|^2/m_{\tilde t}^2$.   In order to satisfy the
LEP bound, one needs either
\begin{enumerate}
\item  $m_{\tilde t} \geq 1$ TeV,  or
\item  $X_t \approx \pm \sqrt{6}$.
\end{enumerate}
The second possibility is known as the large mixing angle limit.
Very special SUSY breaking scenarios are needed to obtain this. The
first possibility is to have a heavy stop. If we now make the
reasonable assumption that all scalar masses at the high energy
scale are of the same order, we run into trouble. Consider the tree
level Z mass given by (for moderate values of $\tan\beta \geq 5$) by
\begin{eqnarray}
\frac{M_Z^2}{2} \approx - m_{H_u}^2(M_Z) - \mu^2 .
\end{eqnarray}
Then for $- m_{H_u}^2(M_Z) \approx  m_{\tilde t}^2 \approx  {\cal
O}$(1 TeV), we need to fine-tune $\mu^2 = m_{\tilde t}^2 (1 -
\epsilon)$ such that
\begin{eqnarray}
\epsilon \approx \frac{M_Z^2}{m_{\tilde t}^2} \leq 10^{-2}.
\end{eqnarray}

This argument can be made a bit more rigorous by defining the SUSY
breaking parameters at the GUT scale and then using the
renormalization group equations to find (for $\tan\beta = 10$)
\begin{eqnarray}
M_Z^2 & = & - 1.9 \mu^2 + 5.9 M_3^2 - 1.2 m_{H_u}^2 + 1.5 m_{\tilde
t}^2 \\
& & - 0.8 A_t M_3 + 0.2 A_t^2 + \dots .  \nonumber
\end{eqnarray}
where the parameters on the RHS are evaluated at $M_{GUT}$. Thus we
see that either way we need to fine-tune parameters by {\cal O}
($10^{-2} - 10^{-3}$).   Of course, this fine-tuning should be
compared to one part in $10^{32}$ in the Standard Model.  So perhaps
it is NOT a huge problem.   Note also that one can minimize the
amount of fine-tuning by having $M_3 \ll$ 1 TeV, i.e. with a light
gluino.

\subsection{Some suggested solutions to the ``Little" Hierarchy
problem}

\subsubsection{Gauge Messenger model}

\begin{figure}
\includegraphics[width=0.45 \textwidth,height=0.25\textwidth,angle=0,clip]{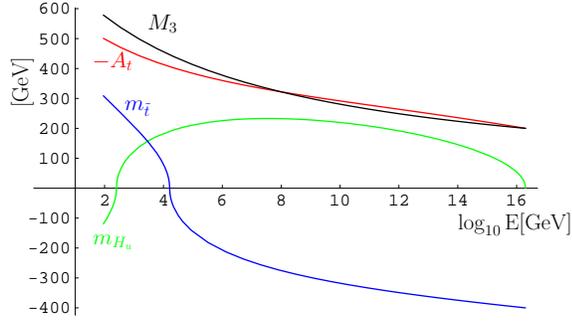}
\caption{Renormalization group running of relevant SSBs for
$\tan\beta = 10$ and GUT scale boundary conditions: $-A_t = M_3 =
200$ GeV, $m_{\tilde t}^2$ = -(400 GeV)$^2$.  Note,  $m_{H_u} \equiv
m_{H_u}^2/\sqrt{|m_{H_u}^2|}$ and $m_{\tilde t} \equiv m_{\tilde
t}^2/\sqrt{|m_{\tilde t}^2|}$. Fig. from Ref.
\cite{Dermisek:2006ey}} \label{fig:1}
\end{figure}

Dermisek and Kim \cite{Dermisek:2006ey} have shown that if one has
soft SUSY breaking boundary conditions which are non-standard, and
in particular, the stop mass squared starts out negative,  then it
is possible to solve the ``little" hierarchy problem.   In Fig.
\ref{fig:1} it is shown that it is possible to obtain $X_t =
\frac{A_t}{m_{\tilde t}}$ large and $M_3$ small, simultaneously.

\begin{figure}
\includegraphics[width=0.45 \textwidth,height=0.25\textwidth,angle=0,clip]{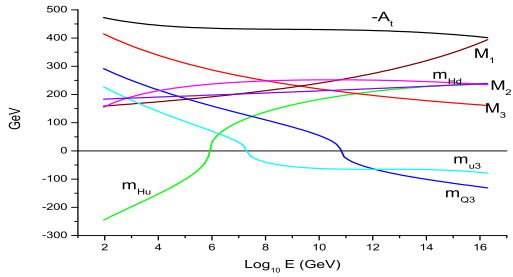}
\caption{Evolution of soft SUSY breaking parameters $-A_t$ (black),
$Q_3, U_3$ mass ([dark, light] blue), $H_u, H_d$ mass (green
magenta) and gaugino masses starting with gauge messenger boundary
conditions at $M_{GUT}$. Note, for example, $m_{H_u} \equiv
m_{H_u}^2/\sqrt{|m_{H_u}^2|}$. Fig. from Ref.
\cite{Dermisek:2006qj}} \label{fig:2}
\end{figure}

Moreover, they consider a scenario with a GUT group SU(5) and an
adjoint $\Sigma$ whose vev breaks  SU(5) to the SM and at the same
time  breaks supersymmetry, such that the effective SUSY breaking
scale $\Lambda = \frac{\alpha_{GUT}}{4 \pi} \left|
\frac{F_\Sigma}{M_{GUT}} \right|$.  This sets the soft SUSY breaking
boundary conditions.  The stop mass squared is naturally negative
and the gaugino masses satisfy: $M_3 = 4 \Lambda$, $M_2 = 6
\Lambda$, $M_1 = 10 \Lambda$.  The find a perfectly acceptable low
energy spectrum with all squarks and sleptons with positive masses
squared.  Moreover, the ``little" hierarchy problem is resolved with
small $M_3$ and significant $X_t$ as seen in Fig. \ref{fig:2} taken
from \cite{Dermisek:2006qj}.
The bottom line for the gauge messenger scenario is
\begin{itemize}
\item  gauge mediation with gauge messengers,
\item   it solves the flavor problem,
\item      5\% to 10 \%  fine tuning for electroweak symmetry breaking,
\item     gravitino  LSP,
\item     stau  NLSP,  but
\item  on the downside, it lacks a UV completion, and
\item   it doesn't address the  $\mu$  problem.
\end{itemize}
\subsubsection{Sweet spot supersymmetry}
Sweet spot supersymmetry (Ibe and Kitano \cite{Ibe:2007km}) combines
gauge and gravity mediation with an effective messenger/ fundamental
scale, $M_{mess} \sim 10^{-3} M_{GUT}$.   The model solves the
flavor problem, since scalar masses are predominantly due to gauge
mediation, with \newline $F/M_{mess} >> F/M_{Pl}$, and not gravity
mediation. It solves the $\mu$ and $B\mu$ problems, a la
Giudice-Masiero.   The authors also discuss a possible UV completion
with GUT breaking and Higgs double-triplet splitting.  The one
downside of this model is that it does not address the ``little"
hierarchy problem.   The gravitino is the LSP with a mass of order 1
GeV.
\subsubsection{Mirage mediation}
\begin{figure}
\includegraphics[width=0.45 \textwidth,height=0.25 \textwidth,angle=0,clip]{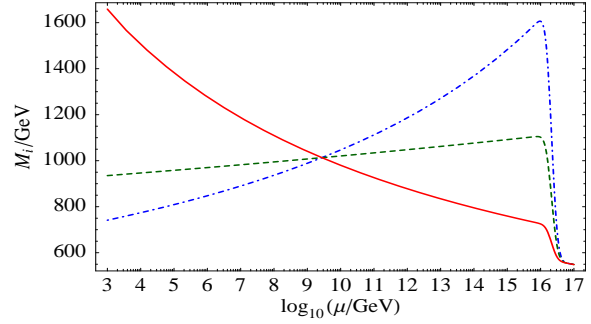}
\caption{For $\alpha = 1$ the gaugino masses meet at a scale of
order $10^9$ GeV. Fig. from Ref. \cite{Lebedev:2005ge}}
\label{fig:3}
\end{figure}

\begin{figure}
\centerline{
\includegraphics[width=0.25\textwidth,height=0.25\textwidth,angle=0,clip]{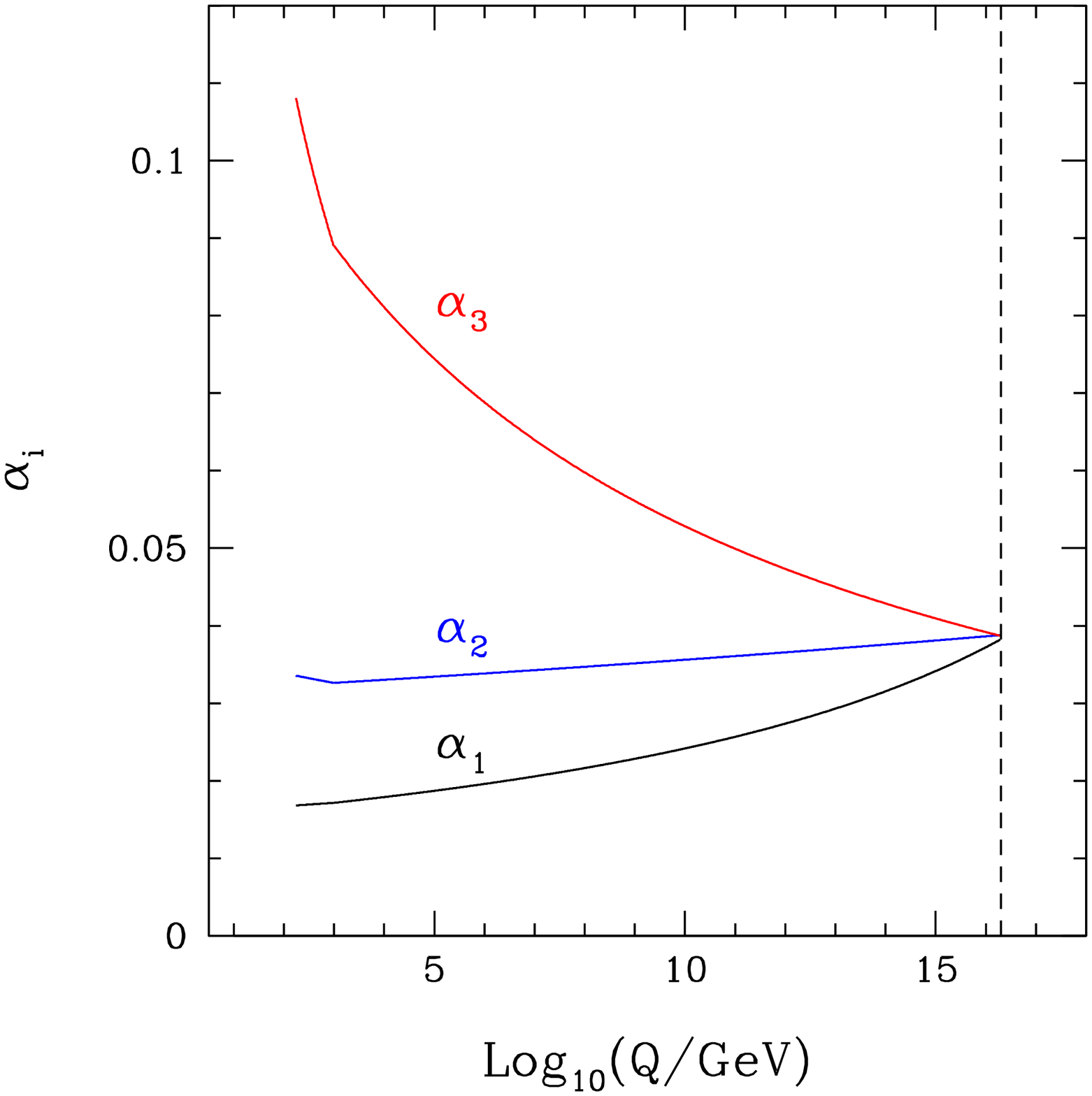}
\includegraphics[width=0.25\textwidth,height=0.25\textwidth,angle=0,clip]{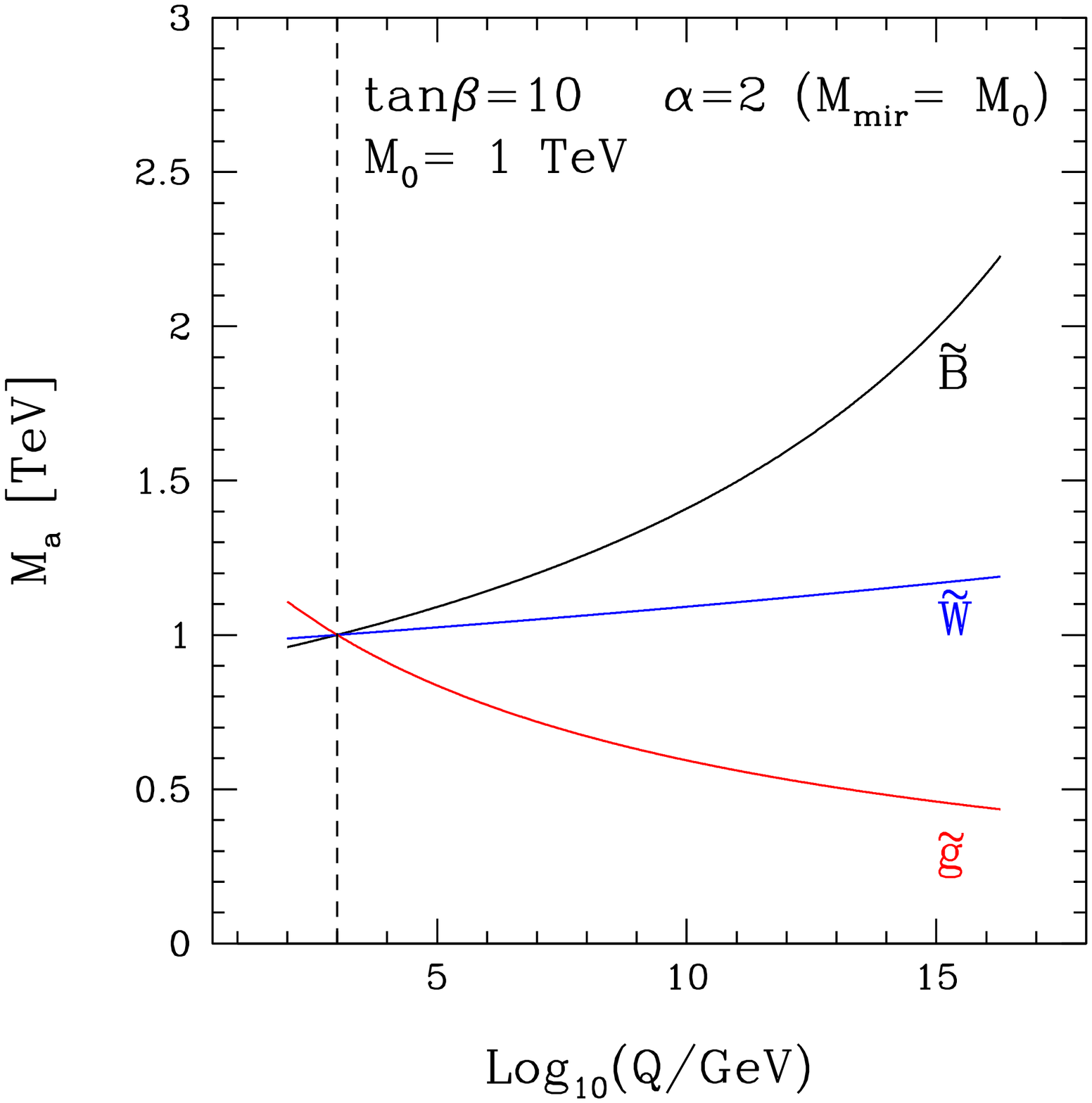}}
\caption{With $\alpha =2$ gaugino masses unify at the weak scale,
while gauge couplings still unify at $M_G \sim 2 \times 10^{16}$
GeV. Fig. from Ref. \cite{Choi:2006xb}} \label{fig:4}
\end{figure}

\begin{figure}
\centerline{
\includegraphics[width=0.25\textwidth,height=0.25\textwidth,angle=0,clip]{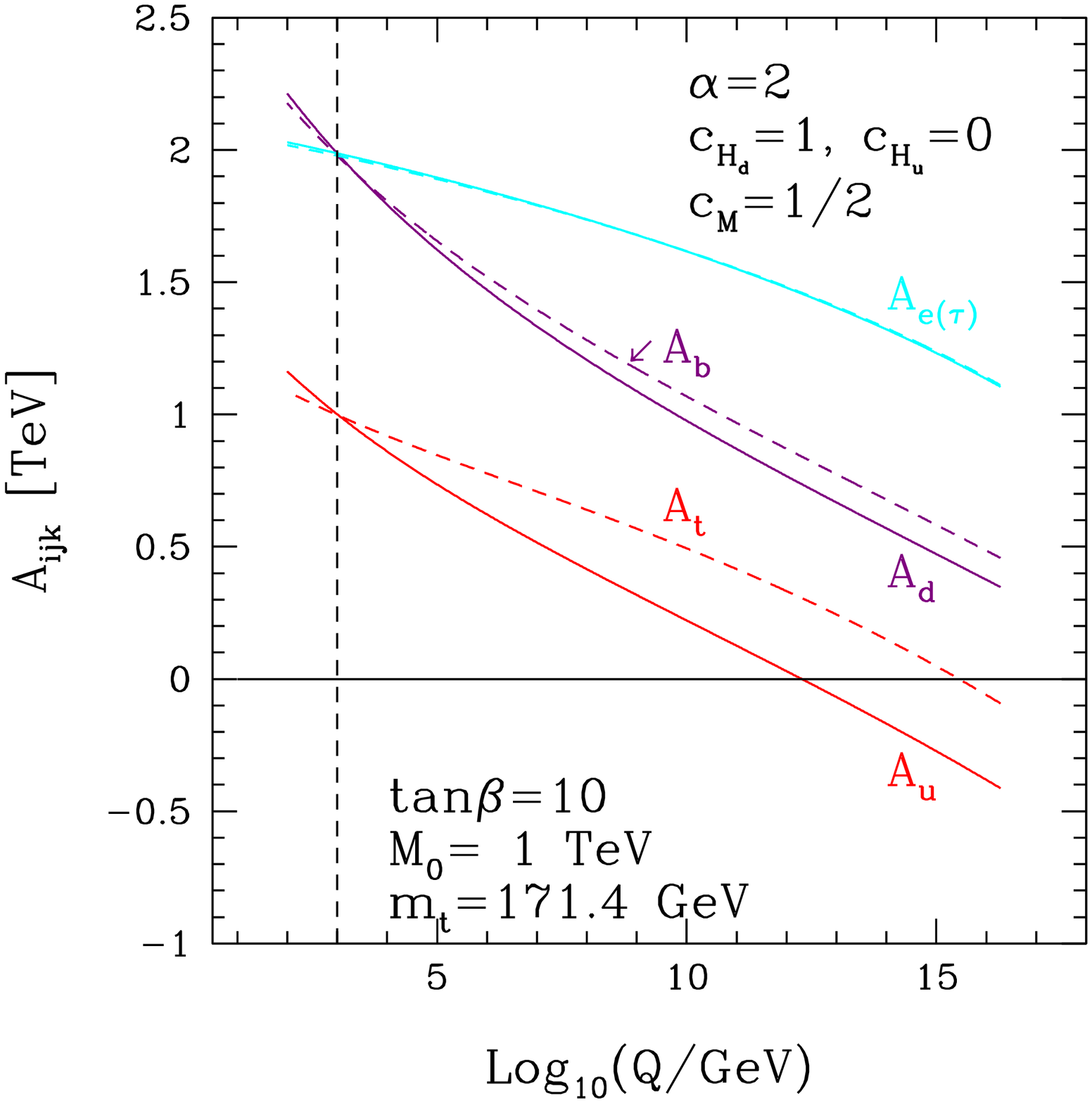}
\includegraphics[width=0.25\textwidth,height=0.25\textwidth,angle=0,clip]{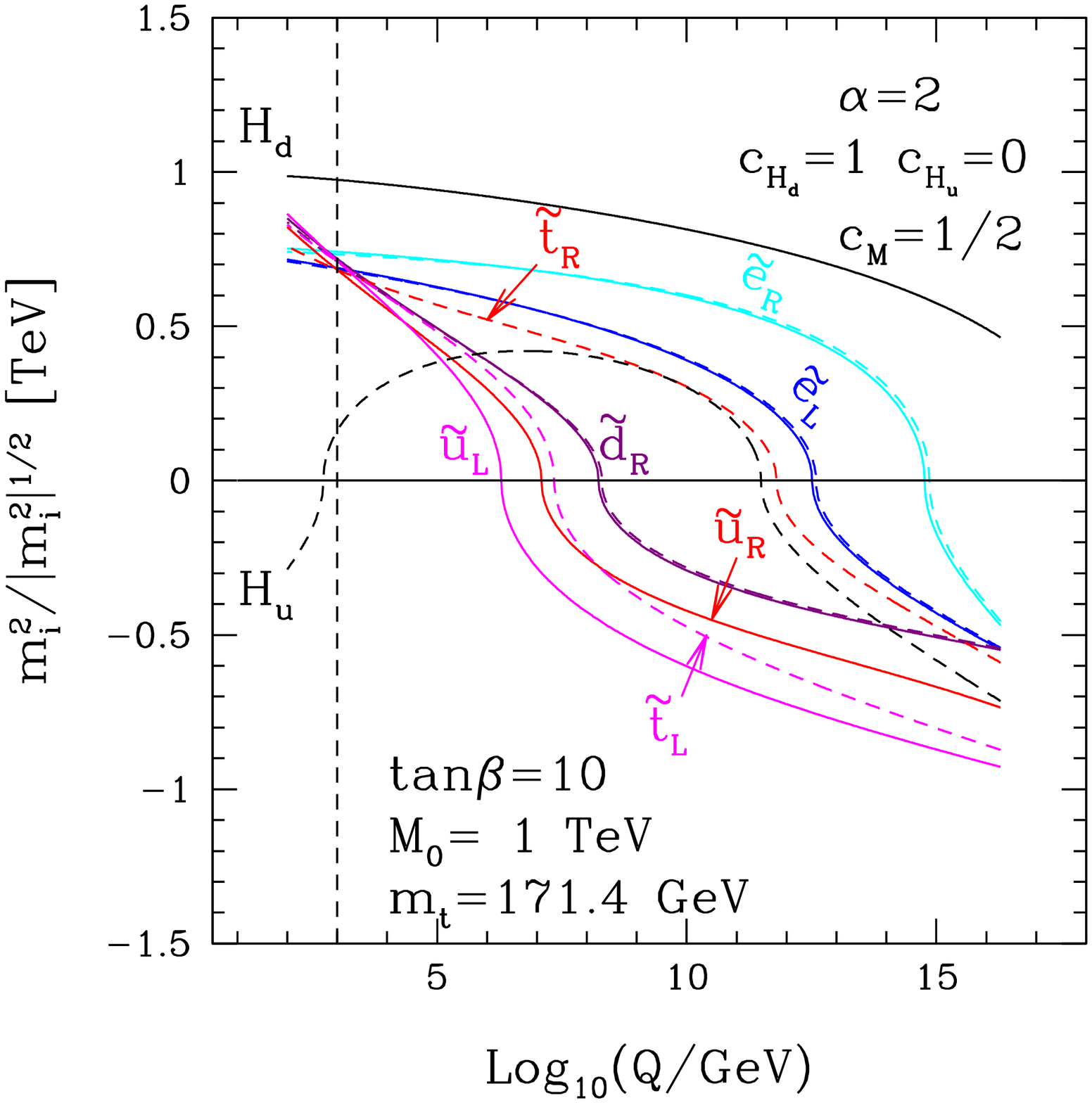}}
\caption{Evolution of soft SUSY breaking parameters in mirage
mediation with $\alpha =2$. Fig. from Ref. \cite{Choi:2006xb}}
\label{fig:5}
\end{figure}

Mirage mediation is based on the work of KKLT \cite{Kachru:2003aw}
demonstrating moduli stabilization and SUSY breaking in Type II
superstrings.   The initial analysis of the consequences of SUSY
breaking in the observable sector is found in Ref.
\cite{Choi:2004sx}. It was realized that the contributions of both
gravity/moduli mediation and anomaly mediation are important
\cite{Endo:2005uy,Choi:2005uz,Falkowski:2005ck,LoaizaBrito:2005fa,Lebedev:2006qc}.
If one defines the ratio $\alpha$ = anomaly : Modulus  SUSY
breaking, then one finds for $\alpha = 1$ that the gaugino masses
unify at a ``mirage" scale of order 10$^9$ GeV (see Fig.
\ref{fig:3}), while gauge couplings continue to unify at the GUT
scale, $M_G \sim 2 \times 10^{16}$ GeV. As discussed in Refs.
\cite{Choi:2005hd,Lebedev:2005ge,Kitano:2005wc,Choi:2006xb}, with a
choice of $\alpha =2$ one can have the gaugino masses unify at the
electroweak scale (see Fig. \ref{fig:4}). Moreover with light
gauginos and sfermions one can now ameliorate the ``little"
hierarchy problem (see Fig. \ref{fig:5}).

The bottom line is that this model has a heavy gravitino with
\begin{eqnarray}  m_{3/2} & \approx & \log(M_{Pl}/m_{3/2}) m_{soft}
\\
& \approx & 4 \pi^2 m_{soft} \approx 10 \;\; {\rm TeV}. \nonumber
\end{eqnarray} In addition \begin{itemize}
\item  it preserves gauge coupling unification at $M_{GUT} \sim 10^{16}$
GeV ;
\item   gaugino masses unify, BUT  typically  below the GUT scale;
\item  it cures the negative slepton mass squared problem of  pure anomaly
mediation; \item and it can  address the "little" hierarchy problem;
\item On the downside,  the solution to the SUSY flavor problem is model
dependent, i.e. it depends on the supposition that there exists a
model in which all fermions reside on a $D7$ brane.
\end{itemize}

\subsection{Gaugino Code}

Choi and Nilles \cite{Choi:2007ka}, after analyzing several
different SUSY breaking mechanisms, suggest that gauginos are a
sensitive window onto the fundamental SUSY breaking mechanism. In
particular,  they find the following gaugino spectra -
\begin{itemize}
\item  mSugra and Gauge mediated SUSY breaking \subitem $M_1 : M_2 : M_3
\cong 1 : 2 : 6 \cong g_1^2 : g_2^2 : g_3^2$;
\item  anomaly pattern \subitem $M_1 : M_2 : M_3
\cong 3.3 : 1 : 9 $;
\item mirage pattern \subitem $M_1 : M_2 : M_3
\cong 1 : 1 : 1$  for $\alpha=2$,
 \subitem $M_1 : M_2 : M_3
\cong 1 : 1.3 : 2.5$  for $\alpha \cong 1$;
\item gauge messenger pattern \subitem $M_1 : M_2 : M_3
\cong 1 : 1.1 : 2$.
\end{itemize}

\section{Higgs ``portal" on physics beyond the MSSM}

There are several indicators that the Higgs bosons are very special.
In SUSY theories, the ``little" hierarchy problem,  the $\mu$
problem and fermion masses all point to the Higgs as being the
``portal" onto new physics beyond the Standard Model.  In the
context of non-SUSY theories, this idea has been expressed in Refs.
\cite{Patt:2006fw,O'Connell:2006wi}.  In SUSY theories, this idea
has come up in several different contexts.  For an effective field
theory analysis in SUSY theories, see
\cite{Brignole:2003cm,Dine:2007xi}.

Consider, for example, the next to minimal SUSY model [NMSSM]. In
the NMSSM
\cite{Ellis:1988er,Drees:1988fc,Ellwanger:1993xa,Pandita:1993tg,Franke:1995xn,King:1995ys,Ellwanger:1996gw,Dine:2007xi}
with an additional gauge singlet field $S$ and superpotential of the
form
\begin{eqnarray}  W & = &  S H_u H_d  +  S^3  \\ & \;\; {\rm or} & \nonumber \\
                    & = &  (\mu + \lambda_S \ S) H_u H_d  +
                    \frac{1}{2} M_S  S^2   \nonumber
\end{eqnarray}
it has been shown that the Higgs may be heavier than the LEPII bound
even with a light stop due to an additional shift in mass.

On the other hand, the Higgs may be lighter than the LEPII bound,
BUT it has a new {\em invisible} decay mode into two light CP odd
Higgs bosons, $a$ \cite{Dermisek:2005ar,Dermisek:2007ah}.   If $m_a
< 2 m_b$, then the decay to two bottom quarks is forbidden.  As a
result a light Higgs with mass $\sim 100$ GeV would not have been
seen at LEP.  Moreover, if $m_a > 2 m_\tau$ the decay $h \rightarrow
2 a \rightarrow 4 \tau$ might have been recorded at LEP ( and might
be seen, if only they re-analyze their data)
\cite{Dermisek:2005ar,Dermisek:2007ah}.

\section{MSSM at large $\tan\beta$}

There were two interesting talks in this conference on the subject
of the MSSM at large $\tan\beta \sim 50$ which I would like to
briefly review. The first is a talk by Heinemeyer.  He addresses the
recent CDF data on an excess of Higgs to two $\tau$ events at a mass
of 160 GeV.   Heinemeyer and collaborators \cite{Ellis:2007ss} show
that it is possible to simultaneously fit the CDF data with the
decay of the CP odd Higgs, $A$ with $m_A = 160$ GeV, and have a
light Higgs, $h$ with $m_h = 115$ GeV.   The region of soft SUSY
breaking parameter space which accomplishes this feat has $A_0
\approx - 2 m_0 \sim 2$ TeV (note, their sign convention for $A_0$
differs from others), non-universal Higgs masses and $\tan\beta \sim
50$. It is important to emphasize that at the same time they are
consistent with data on $b \rightarrow s \gamma$ and the bounds on
the decay $B_s \rightarrow \mu^+ \ \mu^-$.

In the second talk by Belyaev it was shown that there are regions of
MSSM parameter space with $\tan\beta \sim 35$ and a very light Higgs
with mass $m_h = 60$ GeV.  The Higgs would have escaped the LEP
bounds since the branching ratio for the decay $h \rightarrow b b$
is suppressed, while the branching ratio for $h \rightarrow \tau
\tau$ is enhanced \cite{Belyaev:2006rf}.

\section{SUSY GUTs}

Supersymmetric grand unified theories explain charge quantization of
quarks and leptons and give hope of explaining, or at least
providing an organizing principle to explain, the hierarchy of
fermion masses. The gauge group SO(10) is very special in this
regard, since one family of quarks and leptons (including a
right-handed neutrino necessary for a See-Saw mechanism) are
contained in the spinor representation, i.e. $$16 : Q, U^c, D^c, L,
E^c, N^c,$$ where I listed the Weyl spinors in one family.  And the
two Higgs doublets needed in the MSSM are contained in the 10
dimensional representation, i.e. $$ 10 :  H_u, H_d, T, T^c,$$ where
$T, T^c$ are color triplet Higgs.  The color triplets must
necessarily have mass of order the GUT scale, while the Higgs
doublets are effectively massless at this scale.

Thus ordinary 4 dimensional SUSY GUTs have three inherent problems.
The first is the doublet-triplet splitting problem as discussed
above. The second is inventing a GUT symmetry breaking sector which
spontaneously breaks the GUT symmetry and leaves only the MSSM
states below the GUT scale.   And the third problem is the
suppression of the nucleon decay rate below the experimental bounds.
None of these problems is insurmountable. However, the examples
which exist in the literature are far from being pretty and it is
even more difficult to imagine them coming from a more fundamental
theory, such as string theory.

\subsection{Fermion masses in SO(10)}
If the standard electroweak Higgs boson is solely contained in a
{\bf 10} dimensional representation and quarks and leptons are in
{\bf 16}s, there is only one Yukawa coupling that is allowed at the
renormalizable level,  i.e.   $$ \lambda 16 \ 10 \  16 $$ with
$$\lambda_t = \lambda_b =\lambda_\tau = \lambda_{\nu_\tau} \equiv
\lambda.$$ We assume that this is only valid for the third family.

Within the context of SUSY SO(10) there are two different versions
in the literature of a so-called ``minimal" SO(10) SUSY model
including all three families. \begin{enumerate} \item
\begin{eqnarray} W \supset 16 \ 10 \ 16 + 16 \ \overline{126} \ 16 +
16 \ 120 \ 16 .
\end{eqnarray}

This version of ``minimal SO(10)" has been discussed by Aulakh,
Babu, Bajc,  Chen, Fukuyama, Mahanthappa,  Mohapatra, Senjanovic,
etc.   It is the minimal {\em renormalizable} SO(10) SUSY model.

\item \begin{eqnarray}  W \supset 16 \ 10 \ 16 + 16 \ 10 \
\frac{45}{M} \ 16 + \cdots \end{eqnarray}

In this version of ``minimal SO(10)" the fermion hierarchy derives
from a hierarchy of effective non-renormalizable operators
suppressed by some fundamental scale, $M$.  This version has been
discussed by Albright, Anderson, Babu, Barr, Barbieri, Berezhiani,
Blazek, Carena, Dermisek, Dimopoulos, Hall, Pati, Raby, Romanino,
Rossi, Starkman, Wagner, Wilczek, Wiesenfeldt, Willenbrock, etc.
\end{enumerate}
Note, only the latter version has a possible UV completion to string
theory.

Consider a particular ``minimal" SO(10)$\times ( D_3 \times U(1)$
family symmetry) model [DR]
\cite{Dermisek:1999vy,Dermisek:2006dc}.\footnote{This is an example
of the second type of minimal SO(10) SUSY model.} This model
predicts Yukawa coupling unification for the third family. The full
set of 3 $\times$ 3 Yukawa matrices is very simple and very much
constrained by symmetry.   It is given by
\begin{eqnarray}
Y_u =&  \left(\begin{array}{ccc}  0  & \epsilon' \ \rho & - \epsilon \ \xi  \\
             - \epsilon' \ \rho &  \tilde \epsilon \ \rho & - \epsilon     \\
       \epsilon \ \xi   & \epsilon & 1 \end{array} \right) \; \lambda &  \\
Y_d =&  \left(\begin{array}{ccc}  0 & \epsilon'  & - \epsilon \ \xi \ \sigma \\
- \epsilon'   &  \tilde \epsilon  & - \epsilon \ \sigma \\
\epsilon \ \xi  & \epsilon & 1 \end{array} \right) \; \lambda & \label{eq:yukawaD3} \nonumber \\
Y_e =&  \left(\begin{array}{ccc}  0  & - \epsilon'  & 3 \ \epsilon \ \xi \\
          \epsilon'  &  3 \ \tilde \epsilon  & 3 \ \epsilon  \\
 - 3 \ \epsilon \ \xi \ \sigma  & - 3 \ \epsilon \ \sigma & 1 \end{array} \right) \; \lambda &
 \nonumber \\
Y_{\nu} =&  \left(\begin{array}{ccc}  0  & - \epsilon' \ \omega & {3 \over 2} \ \epsilon \ \xi \ \omega \\
      \epsilon'  \ \omega &  3 \ \tilde \epsilon \  \omega & {3 \over 2} \ \epsilon \ \omega \\
       - 3 \ \epsilon \ \xi \ \sigma   & - 3 \ \epsilon \ \sigma & 1 \end{array} \right) \; \lambda .
       & \nonumber
 \end{eqnarray}
 where all the arbitrary order one coefficients are explicitly
 listed.  The model fits all fermion masses and mixing angles very well,
 including neutrinos.  Good fits require the
 soft SUSY breaking parameters to satisfy
 \begin{eqnarray}
 - A_0 \approx 2 m_{16} \gg M_{1/2} \sim \mu ,
 \end{eqnarray}
 non-universal Higgs masses and $\tan\beta \approx 50$
 \cite{Blazek:2001sb}.   At first sight,  this is apparently the
 same region of parameter space discussed in the talk by Heinemeyer.
 However, recently I have learned that the sign of $A_0 \;\; ({\rm
 here})$ is opposite to that of Heinemeyer et al.
 \cite{Ellis:2007ss}.\footnote{Private communication, W. Altmannshofer,  D. Guadagnoli, and D.~M.
 Straub.}  So there is no direct comparison between the results.
 Nevertheless,  there has been a recent analysis of this SO(10)
 model whose results were presented at this conference by
 Altmannshofer, see ``Challenging SO(10) SUSY GUTs with family
symmetries through FCNC processes" \cite{Albrecht:2007ii}.   They
perform a global $\chi^2$ analysis of the DR model
\cite{Dermisek:1999vy,Dermisek:2006dc}. The model has a total of 24
parameters at the GUT scale, which must be varied to compare to low
energy data (see Table \ref{tab:parameters}).  This should be
compared to the 27 parameters in the Standard Model or 32 in the
MSSM. Their analysis confirms previous results, but they now extend
the analysis to include b flavor physics such as $b \rightarrow s
\gamma$, $b \rightarrow s l^+ l^-$, $B \rightarrow \tau \nu$, $B -
\bar B$ mixing and $B_s \rightarrow \mu^+ \mu^-$.   The bottom line
is that the model has some difficulty fitting the processes $b
\rightarrow s l^+ l^-$ and $B \rightarrow \tau \nu$.  Good fits
require heavy scalars with mass greater than 10 TeV.   More models
need to be tested as rigorously in order to eventually find a
Standard GUT model.

\begin{table}
\caption{Parameters in the DR model.} \label{tab:parameters}
\begin{tabular}{|lcc|}
\hline
Sector & \# & Parameters \\
\hline \hline
gauge & 3 & $\alpha_G$, $M_G$, $\epsilon_3$, \\
SUSY (GUT scale) & 5 & $m_{16}$, $M_{1/2}$, $A_0$, $m_{H_u}$, $m_{H_d}$, \\
textures & 11 & $\epsilon$, $\epsilon'$, $\lambda$, $\rho$, $\sigma$, $\tilde \epsilon$, $\xi$, \\
neutrino & 3 & $M_{R_1}$, $M_{R_2}$, $M_{R_3}$, \\
SUSY (EW scale) & 2 & $\tan \beta$, $\mu$ \\
\hline \hline
\end{tabular}
\end{table}

\section{UV completion of effective field theories}

Orbifold GUTs in 5 or 6 dimensions can solve some of the problems of
4 dimensional GUTs.   One starts with a GUT in higher dimensions and
then uses boundary conditions at the orbifold fixed points to break
the GUT symmetry, without a complicated GUT symmetry breaking
sector, and also split Higgs doublets and triplets, by projecting
the triplets out of the theory.   In many cases this has the added
effect of eliminating the baryon number violating operators in SUSY
GUTs.   See for example the work of Kawamura; Hall \& Nomura;
Contino, Pilo, Rattazzi \& Trincherini; Altarelli, Feruglio \&
Masina; Dermisek \& Mafi;  H.D. Kim \& Raby;  Asaka, Buchmuller \&
Covi;  Lee; and Hebecker \& March-Russell.   Of course, the inherent
problem with orbifold GUT field theories is that they are not
renormalizable.  Thus they are effective field theories defined
below some cut-off scale $M^*$.   The penultimate UV completion
would be to embed orbifold GUTs into 10 dimensional string theory.
Then $M^* = M_s$, the string scale. The first steps in this program
have already been taken.
\begin{figure}
\includegraphics[width=0.45\textwidth,height=0.25\textwidth,angle=0,clip]{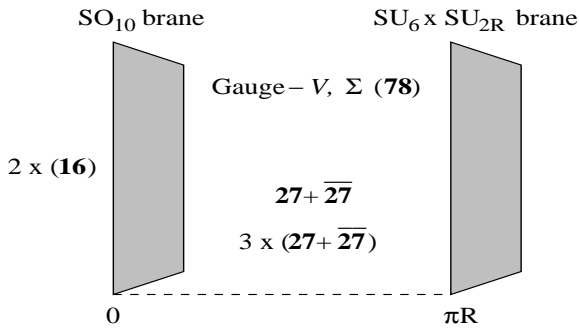}
\caption{E(6) orbifold GUT in $M_4 \times S^1/(Z_2 \times Z'_2)$.
Fig. from Ref. \cite{Kobayashi:2004ya}} \label{fig:6}
\end{figure}

Consider the following E(6) orbifold GUT in 5 dimensions
\cite{Kobayashi:2004ud,Kobayashi:2004ya}. The theory is initially
defined on Minkowski space times a circle. The $Z_{2} \times Z'_{2}$
orbifold leaves a line segment with two boundaries.  At the left
boundary E(6) is broken to SO(10) while at the right boundary it is
broken to SU(6) $\times$ SU(2)$_R$.  The unbroken gauge symmetry is
the intersection of the two, i.e. Pati-Salam - SU(4)$_c \times$
SU(2)$_L \times$ SU(2)$_R$.  The bulk modes include an effective N=2
gauge sector and 4 (27 + 27$^c$ ) hypermultiplets.\footnote{I am
using 4D superfield notation.} In addition there are two families
localized on the SO(10) fixed point (see Fig. \ref{fig:6}).

\begin{figure}
\includegraphics[width=0.45\textwidth,height=0.15\textwidth,angle=0,clip]{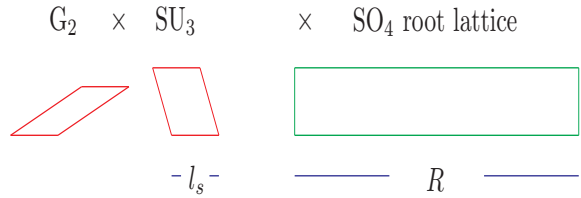}
\caption{$(T^2)^3$ defined in terms of two dimensional planes mod
translations along the vectors of the $G_2 \times SU(3) \times
SO(4)$ root lattices.   Note, we assume one direction is much larger
than the others.} \label{fig:7}
\end{figure}

This model was obtained as the low energy limit of the E(8) $\times$
E(8) heterotic string in 10D
\cite{Kobayashi:2004ud,Kobayashi:2004ya}.   Six of the ten
dimensions are compactified on the product of 3 two tori (Fig.
\ref{fig:7}) defined in terms of the given root lattices.  The tori
are then orbifolded by a $Z_6 = Z_3 \times Z_2$ symmetry with Wilson
lines in the SU(3) and SO(4) torus.  The $Z_6$ symmetry is also
embedded into the E(8) $\times$ E(8) root lattice in terms of a
shift vector $V_6$, consistent with modular invariance constraints.

\begin{figure}
\includegraphics[width=0.45\textwidth,height=0.15\textwidth,angle=0,clip]{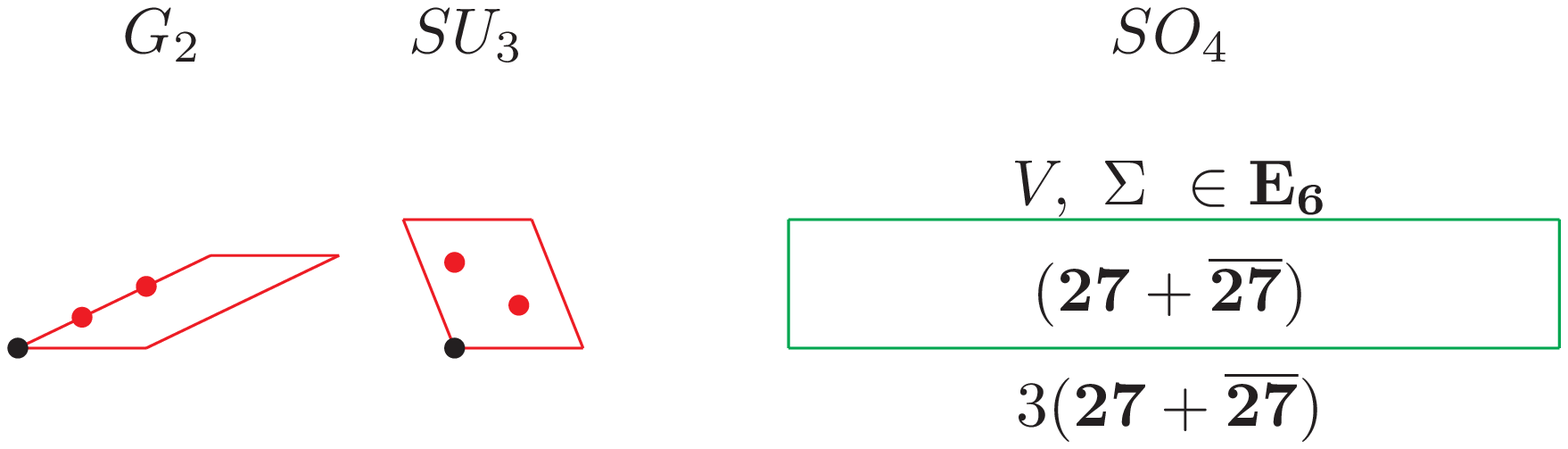}
\caption{These are the massless components of the heterotic string
on $(T^2)^3/Z_3$ plus one Wilson line in the SU(3) torus.  The
massless states come from the untwisted sector, or twisted states
sitting on the pictured fixed points.  All are free to move around
in the SO(4) torus.} \label{fig:8}
\end{figure}

Upon orbifolding the first two tori by $Z_3$ with a Wilson line in
the SU(3) torus one obtains the effective 5 dimensional orbifold
(see Fig. \ref{fig:8}).  The bulk modes, gauge and hypermultiplets,
are the massless string states from the untwisted or $Z_3$ twisted
sectors. All of these string states move freely in Minkowski space
$\times$ the SO(4) torus.  In recent years several groups have
discussed orbifold GUTs from the heterotic string - Kobayashi, Raby
\& Zhang; Forste, Nilles, Vaudrevange \& Wingerter; Buchmuller,
Hamaguchi, Lebedev \& Ratz; JE Kim, JH Kim \& Kyae; and Buchmuller,
Ludeling \& Schmidt.

Heterotic strings in 10 dimensions compactified on a 6 dimensional
compact space have been used to obtain 3 family models with the MSSM
spectrum in 4 dimensions.  See the work of Bouchard, Braun,
Buchmuller, Cleaver, Donagi, Faraggi, Hamaguchi, He,  Kobayashi,
Lebedev, Ludeling, Nanopoulos, Nilles, Ovrut, Pantev, Pokorski,
Raby, Ramos-Sanchez, Ratz, Reinbacher,  Ross, Vaudrevange, Waldram,
Wingerter,  and Zhang.    Also see the talks by Nilles, Kyae,
Luedeling, Lebedev \& Wingerter in this conference.  For a recent
paper on constructing 3 family MSSM models from the heterotic string
see \cite{Lebedev:2007hv} and references therein.

\section{Conclusions}

In recent years, SUSY model building has focused on the ``little"
hierarchy,   $\mu$ and $B \mu$ problems.   There is still no
Standard Model of SUSY breaking. BUT meta-stable vacua appear to be
generic and easy to obtain.  Higgs and gauginos are "portals" on to
new physics beyond the MSSM. Finally, the search for a UV completion
of the MSSM through SUSY GUTs to orbifold GUTs and ultimately to the
heterotic string is now very much in progress.

\subsubsection*{Acknowledgments}
I would like to thank the organizers of SUSY 2007 for a very
engaging meeting.  I also want to dedicate this report to Julius
Wess who is one of the key figures in the discovery and development
of supersymmetry.  Finally, I received partial support from DOE
grant DOE/ER/01545-875.

\end{document}